# PREDICTION OF WAVEFRONTS IN ADAPTIVE OPTICS TO REDUCE SERVO LAG ERRORS USING DATA MINING

Akondi Vyas[1,2], M B Roopashree[1], B R Prasad[1]
[1]Indian Institute of Astrophysics, 2nd Block, Koramangala, Bangalore
[2]Indian Institute of Science, Bangalore
vyas@iiap.res.in, roopashree@iiap.res.in, brp@iiap.res.in

**Abstract:** Servo lag errors in adaptive optics lead to inaccurate compensation of wavefront distortions. An attempt has been made to predict future wavefronts using data mining on wavefronts of the immediate past to reduce these errors. Monte Carlo simulations were performed on experimentally obtained data that closely follows Kolmogorov phase characteristics. An improvement of 6% in wavefront correction is reported after data mining is performed. Data mining is performed in three steps (a) Data cube Segmentation (b) Polynomial Interpolation and (c) Wavefront Estimation. It is important to optimize the segment size that gives best prediction results. Optimization of the best predictable future helps in selecting a suitable exposure time.

## 1. INTRODUCTION

Adaptive Optics (AO) is an indispensable technology in large telescopes to improve the image quality, degraded due to atmospheric disturbances [1]. The major components involved in a simple AO system are - wavefront sensor, wavefront corrector and control algorithm. A wavefront sensor is a device that helps in determining the shape of the incoming beam. Wavefront corrector is a phase distortion compensation tool. The control algorithm takes the input from the wavefront sensor and translates the information into command values that can be addressed to the wavefront corrector.

Wavefront sensors are generally integrated with a detecting mechanism. Most wavefront correctors are devices that are electronically driven. Closed loop AO systems run in real time and the frequency bandwidth decides its performance. The frequency bandwidth is controlled by the operation timescales of the control algorithm, the detector (integrated with sensor) response time scales, finite response timescales of corrector and most importantly, the minimum required exposure time [2]. The frequency bandwidth is exposure time limited in the cases where the reference source is significantly dim. The time delay in the response of the individual components of AO and exposure time at low light levels lead to time lag errors, i.e. the wavefront being corrected and the wavefront that is addressed to the corrector do not correspond to the same time instant [3].

## 2. BACKGROUND

The exposure time is decided by looking at the decorrelation time at the telescope site. Decorrelation time, $\tau_{decorr}$ is the time period over which two wavefronts become completely uncorrelated. Using a simple translating wind model for turbulence, generally the decorrelation time is defined as $\tau_{decorr} = d/V_{wind}$ where, d is the telescope aperture and $V_{wind}$ is the average wind velocity integrated along the column above the telescope aperture.

In Hanle site, the decorrelation timescale is ~18 msec [4]. In the case of 1m telescope, using a magnitude 12 natural guide star, it is possible to collect ~1000 photons within an exposure time of 5ms. The delays in the wavefront reconstruction process and generation of actuator command values lead to a finite delay of ~1ms. This is equivalent to compensating a wavefront from the information that was obtained more than 6 msec earlier including response timescales of instruments. Although it is within the decorrelation time, this will introduce significant inaccuracy in the wavefront correction process. A possible partial solution of this problem is to progressively predict wavefronts arriving after a delay time equivalent to the servo time lag. This was done by some authors using frozen in turbulence approximation applied on linear predictor models and artificial neural networks [3, 5, 6].

There are two important parameters that significantly affect the process of progressive prediction in adaptive optics. The number of past wavefronts that can give optimum predictability of the near future is one factor that needs optimization in real time. The best predictable future is another parameter which is needed to be estimated.

Data mining is a knowledge recovery technology from large databases and is used very effectively for predicting trends in data sets [7]. In this paper, a mining procedure was applied on wavefront sensor data to optimize and adapt to best possible performance in real time. In the case of atmospheric AO, these parameters are even more critical since the coherence length and decorrelation timescales are temporal variables. Atmospheric like turbulence was generated in the lab by using a translating compact disk casing. It was shown that the obtained wavefronts distorted by the casing followed Kolmogorov spatial statistics. Series of wavefronts were obtained as illustrated in section 3 and used for testing the proposed prediction algorithm. The prediction algorithm is based on systematic pixel-wise extrapolation of the wavefront data.

## 3. EXPERIMENTAL DATA

A diffractive optical lens based Shack Hartmann Sensor (SHS) was used for wave-front sensing [8]. Fresnel zones were projected on a spatial light modulator to simulate the effect of a SHS, which is an array of equal focal length lenslets that detect local wavefront gradients via the measurement of relative shifts in the spots formed near the focus of the lenses. A compact disk casing was used to produce the effect of atmospheric turbulence following Taylor's "frozen turbulence" hypothesis in the lab. Vector matrix multiply reconstructor based on least square fitting algorithm was implemented on the SHS geometry to reconstruct the wavefronts from slope measurements [9].

### 3.1 Description of the Experiment

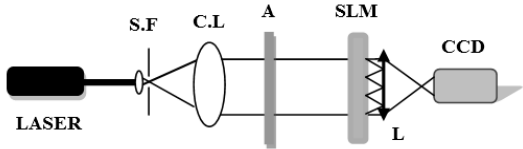

Fig. 1. Experimental setup

The experimental setup is shown in Fig. 1. A 15mW Melles Griot He-Ne laser (633nm) was used as a source of light. The laser beam was cleaned using the Newport compact five axis spatial filter (S.F). The diverging beam was collimated using a collimating lens (C.L.) which is a triplet of focal length 125mm. Aberrations were introduced into the collimated beam using a compact disc casing (A) mounted on a translation stage. A transmitting type nematic liquid crystal based spatial light modulator (SLM-Model LC2002 from Holoeye) was used to project Fresnel zones (12×12).

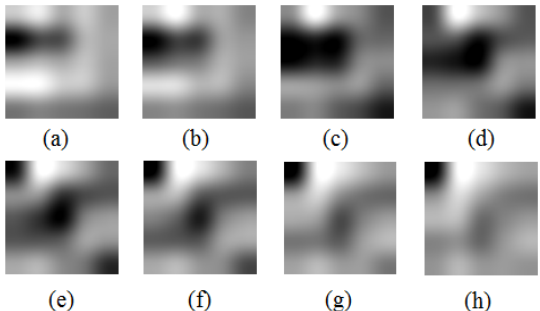

Fig. 2. Experimentally obtained temporally evolving wavefronts (a) to (h)

The focal plane of this SLM based SHS was reimaged onto the CCD (Pulnix-TM-1325CL) using a doublet lens (L) of focal length 150mm. The CD casing was translated perpendicular to the beam in steps of 10μm and the spot pattern as seen on the CCD was recorded at each step. The wavefronts after reconstruction are shown in Fig. 2.

### 3.2 Validating Experimental Data

Most important requirement is that the wavefronts in the time series must become decorrelated for sufficiently large time difference in the sequence of the wavefronts. As a check for this, the correlation coefficient of the wavefront at time t=0 was calculated with wavefronts occurring at later time and is plotted in Fig. 3. for the experimentally obtained data set. It can be seen that the correlation drops to below 30% for the phase screen with index 26. Since on an average any randomly generated phase screens can have ~30% correlation, the 26 phase screen can be considered completely decorrelated from the first phase screen. Associating 18ms decorrelation time to $26^{th}$ phase screen, it can be realized that a servo lag of ~6ms corresponds to $9^{th}$ phase screen which is correlated to the first phase screen by 92.44%.

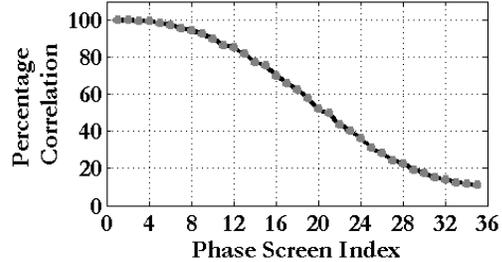

Fig. 3. Correlation of the first wavefront with subsequent wavefronts drops as time increases. Phase screen is a projection of a wavefront on a two dimensional array. Phase screen index is a physical index equivalent to time.

Another check for validity of the data is through a statistical analysis. The structure function D (r) is a common descriptor of atmospheric seeing conditions. It measures the induced phase variance between any two points of the wavefront on the aperture separated by a distance r.

$$D(r) = \langle |\Phi(r - \rho) - \Phi(\rho)|^2 \rangle, \text{averaged over } \rho \quad (1)$$

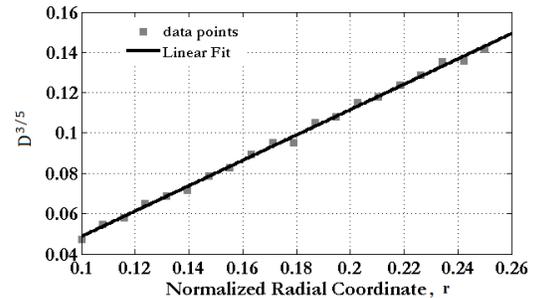

Fig. 4. Experimental data follows Kolmogorov turbulence model

In the case of Kolmogorov model of turbulence, the structure function takes the form D ~ $r^{5/3}$. Hence it is enough to verify that D($r$) for the phase screens calculated using eq. 1 follows the five-third law. D($r$) was computed and averaged over many phase screens and it was observed that $D^{3/5}$ varies linear with $r$ as shown in Fig. 4.

## 4. PREDICTION METHODOLOGY

Data mining is used to predict future trends using time series data through two steps-data segmentation and interpolation [10]. In this analysis, it is convenient to consider phase screens as two dimensional matrices containing pixel values corresponding to phase fluctuations. In the method, the trends in temporal phase fluctuations of individual pixels are studied and predicted. To minimize time and optimize performance data cubes are formed out of the wavefront data.

### 4.1 Formation of Data Cubes

The series of phase screens (two dimensional) reconstructed from the wavefront sensor data were stacked to form a data cube (three dimensional). Over stacking problems were eliminated by removing the older data while new data is being added and hence maintaining a constant cube capacity.

### 4.2 Segmentation Methodology

Segmentation is an important step in prediction process which is associated with piece-wise representation of the time series. Segmentation is performed on a single pixel data that changes over time. A large time series data for an individual pixel is taken and broken into smaller pieces. The last piece is used for extrapolation and hence prediction. Three segmentation techniques namely top-down, bottom-up and sliding windows were implemented. In the sliding windows algorithm, starting from one end of the data set, a segment is grown until it reaches an error threshold and beyond the threshold, a new segment is begun. In the top-down algorithm, the time series is broken recursively until it goes below the threshold error. In contrary, bottom-up algorithm breaks the data with the smallest possible interval and intelligently merges adjacent segments until the error is just below the error threshold. Since top-up algorithm performs faster and is more efficient, it was used in our algorithm for prediction. It can be easily understood that different pixels at different times may need different segment size since some pixels may change more rapidly that others at a same interval of time. This is the reason why this algorithm can perform better than simple linear predictors where segment size is kept constant.

### 4.3 Extrapolation and Wavefront estimation

The last segments from the segmented data sets for individual pixels were used for extrapolation and prediction. Extrapolation can be performed by applying either polynomial interpolation or linear regression over the last segments. After performing extrapolation on data sets for all the pixels, wavefront is formed by suitably clubbing the information from individual pixel predictors. While working with the off-line data, it is possible to compare the predicted wavefronts with the actual ones and evaluate the performance of the adopted methodology.

## 5. RESULTS AND DISCUSSION

Segment size is one of the critical parameters that controls the accuracy of prediction and hence the Strehl ratio. It is therefore needed to optimize this parameter in AO. The best predictable future wavefront helps in optimization of the exposure time and hence the minimum number of photons required for reasonable reconstruction accuracy. These optimization parameters have to be cross checked frequently in real time for better performance of AO predictors. Piecewise linear segmentation (top-down) was used for time series representation of experimentally obtained data. Monte Carlo simulations were performed on the experimentally obtained data cube.

### 4.1 Segment Size

To check the dependence of prediction accuracy on segment size, Monte Carlo simulations were performed and percentage improvement from data predicted using same segment size were averaged and plotted as shown in Fig. 5. The percentage improvement in correlation coefficient, P.I in the correlation coefficient is calculated as follows,

$$P.I = \langle \left(\frac{C_P - C_A}{1 - C_A}\right) \times 100 \rangle$$

(2)

Where, $C_A$ represents the actual correlation coefficient of the first phase screen with the subsequent phase screens and $C_P$ represents the correlation of the first phase screen with the subsequent predicted screens. (1-$C_A$) factor in the denominator brings P.I down when the correlation drops significantly. The improvement in the correlation after prediction is shown in Fig. 5 for a single data set and changing segment size. It was observed that the optimum segment size varies with the data set and most generally it is 3, 4 or 5.

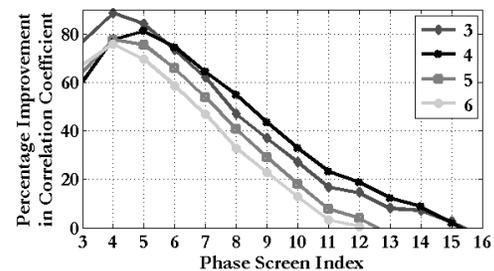

Fig. 5. Performance of the algorithm at different segment sizes

### 4.2 Best Predictable Phase Screen

If the optimum time delay which gives best prediction accuracy is known, the exposure time can be fixed accordingly. It was observed that the best predictable phase screen varies for different data cubes as shown in Fig. 6.

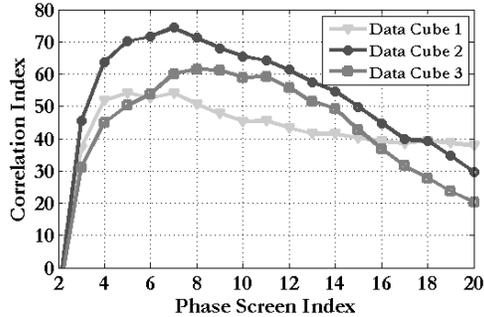

Fig. 6. Best predictable phase screen

It is evident from the simulations that the improvement depends on segmentation, decorrelation timescales, servo lag error timescales and the fitting methodology. Very large and too small segmentation sizes led to worse prediction. Although it is time consuming, linear regression is better than linear interpolation. The study of best predictable phase screens using data mining for certain experimental parameters helps astronomers to optimize the exposure time and design the device parameters in adaptive optical imaging systems. On an average, an improvement of 6% (from 89% to 95%) in the correlation between the actual wavefront and the wavefront addressed to the corrector was observed after prediction when the servo lag time is one-third of the total decorrelation time as shown in Fig. 7. Errors due to servo lags become even more critical when the wave-front reconstruction algorithms are inaccurate and time consuming.

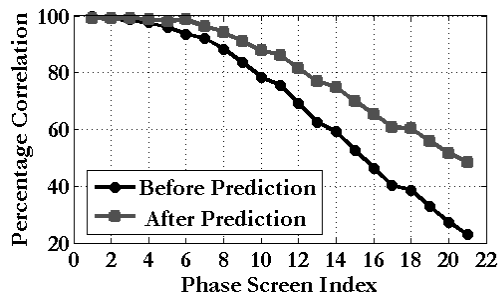

Fig. 7. Percentage Improvement in correlation before and after prediction algorithm is used